\documentclass[twoside,11pt]{article}

\usepackage{graphicx}

\usepackage{amssymb}
\usepackage{amsmath}
\usepackage{amsfonts}
\usepackage{wasysym}
\usepackage{revsymb}
\usepackage{amsbsy}
\usepackage{bm}

\usepackage{titlesec}

\usepackage{cite}

\titleformat*{\section}{\bfseries}
\titleformat*{\subsection}{\bfseries}

\begin{document}           % End of preamble and beginning of text.

\title{\Large 
The Hubble tension from the standpoint of quantum cosmology}
\author{V.E. Kuzmichev, V.V. Kuzmichev\\[0.5cm]
\itshape Bogolyubov Institute for Theoretical Physics,\\
\itshape National Academy of Sciences of Ukraine, Kyiv, 03143 Ukraine}

\date{}

\maketitle

\begin{abstract}
The Hubble tension is analyzed in the framework of quantum cosmological approach. It is found that there arises a new summand in the expression for the 
total energy density stipulated by the quantum Bohm potential. This additional energy density acts similarly to a stiff matter component, modifying the 
expansion history of the early universe and decaying with scale factor $a$ as  $a^{-6}$, faster than radiation, in late universe. Taking account of this 
matter-energy component of quantum nature can, in principle, eliminate a discrepancy between the direct late time model-independent measurements of the 
Hubble constant and its indirect model dependent estimates. The considered model allows one to extend the standard cosmology to quantum sector.
\end{abstract}


\section{Introduction}\label{sec:1}
The Hubble expansion rate (or the Hubble constant) is a cornerstone of the modern cosmology based on the idea of non-stationary
expanding universe. Its value $H_{0}$ in the modern epoch can be extracted from astronomical data by means of independent analyses within
the framework of different model concepts. There is a discrepancy between the Hubble constant estimates obtained from classical distance 
ladder approaches using Cepheid-calibrated SNe Ia and those derived from CMB data that take into account the standard $\Lambda$CDM model,
$H_{0\, \mbox{\tiny SNe}} = 73.04 \pm 1.04$ km/s/Mpc \cite{R22} and $H_{0\, \mbox{\tiny CMB}} = 67.4 \pm 0.5$ km/s/Mpc \cite{Planck18}, respectively. 
The values presented here are somewhat illustrative, because there are not only these two values, but actually two sets of measurements, the direct late time 
$\Lambda$CDM-independent measurements and the indirect model dependent estimates at early times  \cite{KR22,CI22}. It might be argued that the 
disagreement in them, also known as ``Hubble tension'', can hardly be explained by any significant systematic errors in determination of cosmological 
parameters \cite{Kn19,DS21,DS22}. Therefore, this circumstance provides a good reason to search for ``new physics'' beyond the standard $\Lambda$CDM 
model. Another tension problem intensively pursued in current cosmology is the $S_{8}$ tension which is not as statistically significant as Hubble tension  
\cite{CI22}. It refers to the amplitude of matter density fluctuations at the present epoch being smaller than predicted in $\Lambda$CDM cosmology, also based on the CMB measurements (see, e.g., Refs.~\cite{J22,CH23,L23,A23}).

An introduction of a new element into the standard cosmological model may be a key allowing to relieve this, the ``third'' from the historical point of view 
\cite{KR22}, Hubble tension. The possible solution to the Hubble constant problem lies in modification of scale factor evolution before recombination. 
In particular, there is a subclass of Early Dark Energy (EDE) models that postulate the existence of an additional energy density component that peaks before 
recombination and decays more rapidly than radiation shortly afterwards. The EDE models meet with a number of issues, which include, among other things, 
some level of fine-tuning and slight worsening of the $S_{8}$ tension \cite{Ka16,CH20,V21,L22}. The detailed overview of modern approaches to address the 
cosmological tensions problems can be found in Refs.~\cite{CI22,R21,DS23,V23,C20,C22,Ny23}.

In the present note, we point out the feasible solution to the Hubble tension based on consecutive consideration of the evolution of the early universe
in the epoch, when quantum corrections to the Einstein--Friedmann equation are significant and cannot be neglected. Usually, it is accepted that quantum 
theory is of relevance only in the microscopic regime. However, this is not always true. If we agree that the universe as a whole is of quantum nature, there 
may exist situations, where a classical behavior breaks down and its quantum nature becomes apparent (see, e.g., Ref.~\cite{Ki08}). We demonstrate that 
quantum effects generate an additional energy density which decays with scale factor $a$ as  $a^{-6}$, i.e. faster than radiation, and affects the evolution of 
the early universe.

The paper is organized as follows. In Sect.~\ref{sec:2}, the basic concepts and the equations of the quantum cosmological model are introduced. 
The expression for the Hubble expansion rate with the correction term produced by the Bohm potential is obtained in Sect.~\ref{sec:3}.
In Sect.~\ref{sec:4}, it is shown that the quantum addition to the energy density can be considered as a stiff matter (fluid) with the dynamical coupling constant.
Overview of several examples of how matter components with the stiff equation of state can appear in the cosmological models without involving quantum 
cosmology is given in Sect.~\ref{sec:5}. In Sect.~\ref{sec:6}, numerical estimations of the dynamical coupling constant determining the stiff matter component 
before recombination are given in the case when the additional stiff energy density makes about ten-percent contribution to the total energy density providing, 
in particular, a way to solve or alleviate the Hubble tension problem. Finally, in Sect.~\ref{sec:7}, the main results are summarized. The possible generalization 
of the standard $\Lambda$CDM model to quantum sector is discussed here.
 
It is convenient to define the variables making them dimensionless. We use the modified Planck system of units. The 
$l_{p} = \sqrt{2 G \hbar/(3 \pi c^{3})} = 0.7444 \times 10^{-33}$ cm
is taken as a unit of length, the $\rho_{p} = 3 c^{4}/(8 \pi G l_{p}^{2}) = 1.628  \times 10^{117}$ GeV/cm$^{3}$ is a unit of energy density and so on.

\section{Equations of the model}\label{sec:2}
We consider the homogeneous and isotropic quantum cosmological system (universe) whose geometry is determined by the Robertson--Walker line element
with the cosmic scale factor $a$. It is assumed that such a universe is originally filled with a uniform scalar field $\phi$ and a reference perfect fluid. 
After averaging with respect to appropriate quantum $k$-states $u_{k}(\phi)$, the scalar field turns into the effective barotropic fluid \cite{K13,K16} with the 
energy density $\rho_{m} = 2 M/ a^{3}$, where $M$ is an amount of non-relativistic matter which may include dark matter in the non-relativistic sector. In the 
general case, the mass $M$ may depend both on the quantum number $k$ of the wave function of matter $u_{k}(\phi)$ in the representation, in which the 
Hamiltonian of the uniform scalar field $\phi$ is diagonalized, and on the cosmic scale factor $a$, $M \equiv M_{k}(a)$. The value $a^{3}/2$ plays a role of a 
proper volume made dimensionless.

The reference perfect fluid defines a matter reference frame \cite{Ku91,BM96,K02,K08} and is taken in the form of relativistic matter (radiation) with the energy
density $\rho_{\gamma} = E/a^{4}$, where $E = const$ (in natural physical units, it is proportional to the conversion constant $\hbar c$). The density
$\rho_{\gamma}$ may include dark matter in the relativistic sector. Dark energy is simulated by the cosmological constant $\Lambda$.

Such a model universe is described by the state vector \cite{K13,K08}
\begin{equation}\label{1}
\Psi (a,\phi;T) = e^{\frac{i}{2} ET} \sum_{k} u_{k}(\phi) \psi_{k}(a),
\end{equation}
where $T$ is the time variable (conformal time) connected with the proper time $t$ by the differential equation $dt = a dT$, and $dT = N d \eta$,
$N$ is the lapse function whose choice is arbitrary \cite{ADM62}, $\eta$ is the ``arc time'' which coincides with $T$ for $N = 1$ (cf. 
Ref.~\cite{MTW73,LL75}). In general case, the function $N$ serves as a Lagrange multiplier in the Hamiltonian constraint formalism and
it should be taken into account in an appropriate way.

The wave function $\psi_{k}(a)$ satisfies the equation:
\begin{equation}\label{2}
\left [- \partial_{a}^{2} - a^{4} \rho(a)\right ] \psi (a)  = 0,
\end{equation}
where
\begin{equation}\label{3}
\rho(a) = \frac{2 M}{a^{3}} + \frac{E}{a^{4}} + \frac{\Lambda}{3} - \frac{\kappa}{a^{2}}
\end{equation}
is the total energy density. Here the subscript $k$ is omitted and the curvature term $- \kappa/a^{2}$ with $\kappa = +1, 0, -1$ is included into 
the total energy density for convenience.

Formally Eq.~(\ref{2}) coincides with the Wheeler-DeWitt equation \cite{W68,DW62,DW67} for the universe under consideration. However, it contains 
an important difference: the constant $E$ not only enters the energy density (\ref{3}), but also determines the evolution of the universe in conformal time $T$ 
according to Eq.~(\ref{1}). The analogy with the quantum mechanics in the Schr\"{o}dinger--Madelung \cite{Ma27} formulation rather than the Wheeler-DeWitt 
quantum geometrodynamics comes to mind. The latter in principle does not contain a time variable. The presence of the constant $E$ in Eq.~(\ref{3}) is not 
enough. It is required to know what canonical variable should be related to it. In our approach, it is the conformal time $T$, the same as in general relativity.
Let us note that in natural physical units $E$ and $T$ have the dimensions: $[E] = \mbox{Energy} \times \mbox{Length}$, $[T] = \mbox{Radians}$.

\section{The Hubble expansion rate with quantum correction}\label{sec:3}
We look for the solution in the form of a wave propagating along the $a$ direction
\begin{equation}\label{4}
\psi (a)  = |\psi| e^{i S(a)},
\end{equation}
where $|\psi| = const / \sqrt{\partial_{a} S(a)}$ is the amplitude, $S(a)$ is the phase which is assumed to be a real function of the scale factor $a$.
Substituting Eq.~(\ref{4}) into (\ref{2}), we find that $\partial_{a} S(a)$ satisfies the non-linear equation (generalized Hamilton--Jacobi equation),
\begin{equation}\label{5}
\left(\partial_{a} S(a) \right)^{2} = a^{4} \left(\rho + \rho_{B} \right),
\end{equation}
where the term
\begin{equation}\label{6}
\rho_{B} = \frac{Q_{B}}{a^{4}}
\end{equation}
is known in quantum mechanics as the energy density produced by the Bohm potential \cite{B52},
\begin{equation}\label{7}
Q_{B} = \frac{3}{4} \left(\frac{\partial_{a}^{2} S}{\partial_{a} S} \right)^{2} - \frac{1}{2} \frac{\partial_{a}^{3} S}{\partial_{a} S}. 
\end{equation}

In order to pass from Eq.~(\ref{5}) to the equation for the Hubble constant, the time variable $T$ should be restored. Following Dirac \cite{D58} and
considering the wave function $\psi (a)$ as immovable vector of the Heisenberg representation, we obtain the equation of motion \cite{K13}
\begin{equation}\label{8}
\langle \psi | -i \partial_{a} | \psi \rangle = \langle \psi | - \frac{da}{dT} | \psi \rangle,
\end{equation}
which connects the quantum-mechanical momentum $-i \partial_{a}$ with the velocity $da / dT$ of general relativity. The simple substitution
 of Eq.~(\ref{4}) into (\ref{8}) gives the expression
\begin{equation}\label{9}
\partial_{a} S + \frac{i}{2} \frac{\partial_{a}^{2} S}{\partial_{a} S} = - \frac{da}{dT}.
\end{equation}
From Eq.~(\ref{9}), it follows that time $T$ should be complex for real $a$ and $S(a)$. This means that for complete description of the dynamics of the 
universe one has to take into account that the universe can transit from the region near initial singularity with sub-Planck scales, where the interval has the 
Euclidean signature and time is complex, to the region with the Lorentzian signature metric and real-valued physical variables (for discussion see 
Refs.~\cite{K09,HH83,HP00}). Let us notice here that such a picture of change in spacetime geometry during the transition of the universe from the region near 
initial singularity into the region of real physical scales can be interpreted as the spontaneous nucleation of the expanding universe from the initial
singularity point.

Taking into account that time $T$ can be complex, $T = T_{R} + i T_{I}$, where $T_{R}$ is time in real  physical region, whereas $T_{I}$ is imaginary
time component on sub-Planck scales, we have
\begin{equation}\label{10}
\frac{da}{dT} = \frac{\partial a}{\partial T_{R}} - i \frac{\partial a}{\partial T_{I}},
\end{equation}
where, according to Eq.~(\ref{9}),
\begin{equation}\label{11}
- \frac{\partial a}{\partial T_{R}} = \partial_{a} S, \quad \frac{\partial a}{\partial T_{I}} = \frac{1}{2} \frac{\partial_{a}^{2} S}{\partial_{a} S},
\end{equation}
and Eq.~(\ref{5}) can be rewritten as
\begin{equation}\label{12}
\left(\frac{\partial a}{ \partial T_{R}} \right)^{2} = a^{4} \rho + Q_{B}.
\end{equation}
This equation can be derived directly from Eq.~(\ref{5}). To do this, one should square the left- and right-hand sides of Eq.~(\ref{9}) and use 
Eqs.~(\ref{10}) and (\ref{11}).

Since the Hubble expansion rate $H$ has meaning only in the region of real physical values, then defining it in accordance with general relativity,
\begin{equation}\label{13}
H = \frac{1}{a^{2}} \frac{\partial a}{\partial T_{R}} = - \frac{\partial_{a} S}{a^{2}},
\end{equation}
we obtain the equation
\begin{equation}\label{14}
H^{2} = \rho + \rho_{B},
\end{equation}
which contains an additional energy density (\ref{6}) produced by the quantum potential (\ref{7}).

Using Eqs.~(\ref{7}) and (\ref{13}), one can rewrite Eq.~(\ref{6}) in the form
\begin{equation}\label{141}
\rho_{B} = \frac{2}{a^{6}} + \frac{1}{a^{5}} \frac{\partial_{a} H}{H} + \frac{1}{a^{4}} \left[\frac{3}{4} \left(\frac{\partial_{a} H}{H} \right)^{2} -
\frac{1}{2} \frac{\partial_{a}^{2} H}{H}\right].
\end{equation}
Then Eq.~(\ref{14}) can be considered as a nonlinear equation for the Hubble parameter $H$. The expression (\ref{141}) can be represented 
as a nonlinear equation for the quantum addition to energy density
\begin{equation}\label{142}
\rho_{B} = \frac{2}{a^{6}} + \frac{1}{2 a^{5}} \frac{\partial_{a} (\rho + \rho_{B}) }{\rho + \rho_{B}} + \frac{1}{4 a^{4}} \left[\frac{5}{4} 
\left(\frac{\partial_{a} (\rho + \rho_{B})}{\rho + \rho_{B}} \right)^{2} - \frac{\partial_{a}^{2} (\rho + \rho_{B})}{\rho + \rho_{B}}\right].
\end{equation}

Since the Bohm potential (\ref{7}) is small in comparison with the term $a^{4} \rho$ in Eq.~(\ref{5}) (the potential $Q_{B}$ written in the natural physical units
is proportional to $\hbar^{2}$ \cite{K13,K09}), then Eq.~(\ref{142}) can be solved using perturbation theory.

From dimensional reasons, it follows that both representations of $\rho_{B}$, (\ref{141}) and (\ref{142}), lead to the same dependence on $a$,
$\rho_{B} \sim a^{-6}$, up to a dimensionless factor (we denote it as $\gamma$) which can be calculated in an explicit form.

\section{Quantum potential in a single dominant matter field model}\label{sec:4}
Let us find explicit dependences of $Q_{B}$ (\ref{7}) and $\rho_{B}$ (\ref{6}) on $a$ in the model of a single dominant matter field. 
According to general relativity, for the most common matter sources, the dependence of $H$ on proper time $t$ 
has a simple form: $H = \frac{\alpha}{t}$, where $\alpha$ is a parameter defined by a type of matter (e.g., $\alpha = \frac{1}{2}$ for radiation, 
$\alpha = \frac{2}{3}$ for non-relativistic matter and so on). Then from the differential equation 
$\frac{\dot{a}}{a} \equiv \frac{1}{a} \frac{da}{dt} = \frac{\alpha}{t}$, it follows the solution 
$a = \beta t^{\alpha}$, where $\beta$ is an integration constant. From Eqs.~(\ref{11}) and (\ref{13}), we find that 
$\partial_{a} S = - \alpha \beta^{1/\alpha} a^{2 - 1/\alpha}$. Calculating higher derivatives and substituting them in Eq.~(\ref{7}), we get
\begin{equation}\label{15}
Q_{B} = \frac{\gamma}{a^{2}},
\end{equation}
where the numerator
\begin{equation}\label{16}
\gamma = \frac{(2 \alpha - 1) (4 \alpha - 1)}{4\alpha^{2}}
\end{equation}
does not depend on $a$.

For the universe expanding exponentially, $a = a(0) e^{\sqrt{\rho_{v}} t}$, 
where $\rho_{v}$ is some constant (e.g., $\rho_{v} = \frac{\Lambda}{3}$), the Hubble constant is $H = \sqrt{\rho_{v}}$ and the momentum
equals to $\partial_{a} S = - \sqrt{\rho_{v}} a^{2}$. For such a universe, the quantum potential has a form
\begin{equation}\label{17}
Q_{B} = \frac{2}{a^{2}}.
\end{equation}
It is worth noting that Eq.~(\ref{17}) follows from (\ref{15}) in the limit $\alpha \rightarrow \infty$.

Thus the quantum addition to the energy density is equal to
\begin{equation}\label{18}
\rho_{B} = \frac{\gamma}{a^{6}}.
\end{equation}
According to Eq.~(\ref{16}), the energy density $\rho_{B}$ can be negative, when $\frac{1}{4} < \alpha < \frac{1}{2}$, positive for
$\alpha > \frac{1}{2}$ and $\alpha < \frac{1}{4}$, or vanish at $\alpha = \frac{1}{2}$ and $\alpha = \frac{1}{4}$. Let us note that if a single dominant matter
field is radiation ($\alpha = \frac{1}{2}$) or the exotic ekpyrotic matter ($\alpha = \frac{1}{4}$), the latter corresponds to the equation of state parameter
$w = \frac{p}{\rho} = \frac{5}{3} > 1$, then no additional quantum energy density $\rho_{B}$ is produced in such universes.
For $\alpha \ll 1$, the numerator $\gamma$ in the energy density expression (\ref{18}) becomes large, $\gamma \gg 1$.

%Fig.~1
\begin{figure*}% figure* for wide figure, [h] [!] to change the placement
\centering
\includegraphics[width=10cm]{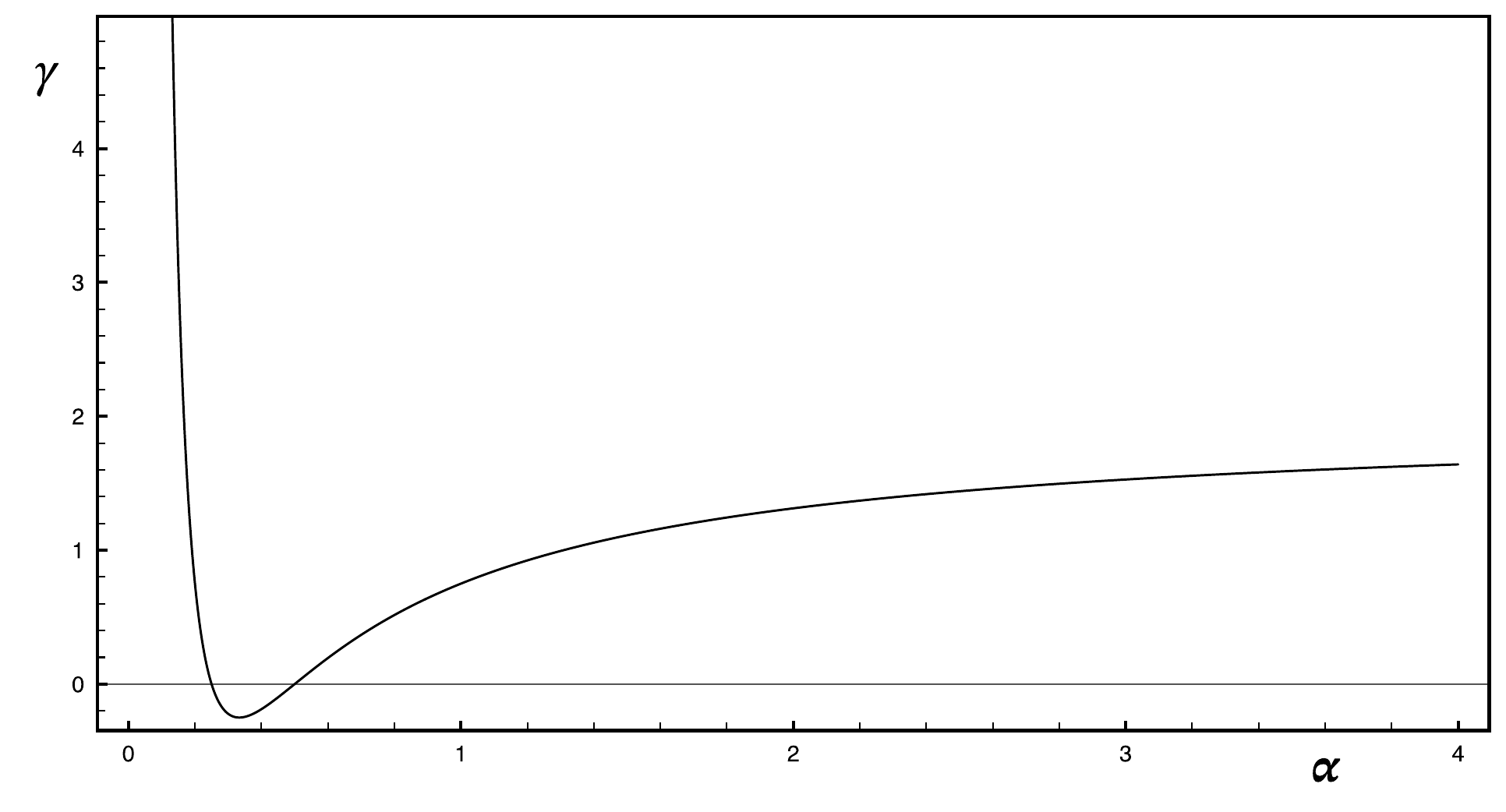}
\caption{The factor $\gamma$ (\ref{16}) as a function on $\alpha$.}
\label{fig1}
\end{figure*} 

The factor $\gamma$ (\ref{16}), considered as a function on $\alpha$, is shown in Fig.~\ref{fig1}. 
It has the minimum at $\alpha = \frac{1}{3}$, where $\gamma = - \frac{1}{4}$
and $\rho_{B} = - \frac{1}{4 a^{6}}$ \cite{K09}. The value $\alpha = \frac{1}{3}$ corresponds to the model of a single dominant matter field in the form of a stiff 
fluid with the equation of state $p = \rho$, which generates the quantum addition $\rho_{B} < 0$ in Eq.~(\ref{14}).
Considering the quantum addition as a fluid with the equation of state $p_{B} = w_{B} \rho_{B}$ which decays as $a^{-3(1+w_{B})}$, we find that the additional 
energy density $\rho_{B}$ corresponds to a stiff fluid with $w_{B} = 1$. The negative energy density $\rho_{B}$
corresponds to the negative pressure $p_{B} < 0$. In this case, it means that expanding universe obtains an additional acceleration within the scope of the 
potential $Q_{B}$. The positive energy density $\rho_{B}$ increases the total energy density leading to higher $H$ according to Eq.~(\ref{14}), so that it can, in 
principle, remove a discrepancy between the lower value of the Hubble constant deduced from CMB measurements and galaxy surveys and the higher value 
obtained from measurements of distances and redshifts in the local universe. 

In conventional big bang cosmology, the universe goes through stages in which different matter-energy components dominate. So the universe after the big 
bang passes through the stages of inflation, radiation dominance, matter dominance, and finally, in the modern epoch, dark energy dominance.
In the model under consideration, where the Hubble expansion rate $H = \frac{\alpha}{t}$, it means that the parameter $\alpha$ should change in value 
as the universe evolves. If the cosmological model contains multiple matter-energy components, the parameter $\alpha$ and the corresponding 
equation-of-state (EoS) parameter $w$ (see below) can be considered as some effective quantities that take a wide range of values depending on the model 
under consideration.

Assuming that the universe goes through a stage at which $\dot{a} \ll 1$ and the scale factor remains close to some value,
we find that $\alpha \ll 1$ and $\gamma$ may become very large. As is known, the existence of a ``coasting period'' during which the scale factor $a$ remains
close to the value at which $\dot{a}$ has its minimum is a remarkable feature of the Lema\^{i}tre model. The Eddington--Lema\^{i}tre model is
characterized by an infinitely long ``coasting period'' as the scale factor $a$ asymptotically approaches the value of the static Einstein model \cite{We72}.
Unlike the above models, in the model studied here, the state in which the scale factor  $\dot{a} \ll 1$ can be realized, generally speaking, even if space is not 
positively curved. In the domain of negative values of the factor $\gamma$, there also may be a ``coasting period'' or at least a ``coasting point'' with the scale 
factor $a \approx a_{st}$, where the value 
$a_{st}$ can be found from the equation
\begin{equation}\label{180}
2M a_{st}^{3} + E a_{st}^{2} + \frac{\Lambda}{3} a_{st}^{6} - \kappa a_{st}^{4} + \gamma = 0,
\end{equation}
when $H^{2} (a_{st})= 0$. This will lead to an effective parameter $\alpha \ll 1$ generating an appropriate large value $\gamma \gg 1$ in the next 
step of the universe's evolution.

One can calculate the factor $\gamma$ from Eq.~(\ref{142}) directly. Let us consider a model in which the energy density (\ref{3})
can be represented by a single component in the form of an effective barotropic fluid with the equation of state $p = w \rho$, where $w = const$.
Note that, in quantum cosmology, starting from a model with the scalar field and then averaging the quantum equation of motion over quantum states that 
diagonalize the Hamiltonian of the scalar field, we get the same equation of state \cite{K13}.

In the first approximation of the perturbation theory, a small addition $\rho_{B}$ to the energy density $\rho$ on the right-hand side of Eq.~(\ref{142})
can be neglected. An effective barotropic fluid satisfies the local law of energy conservation in the expanding universe which has a form:
$\partial_{a} \rho = - \frac{3}{a} (p + \rho)$, where the quantum addition $\rho_{B}$ is neglected \cite{K13}. Then, after simple calculations, we get
Eq.~(\ref{18}) with
\begin{equation}\label{181}
\gamma = 2 - \frac{9}{16} (1 + w) (3 - w).
\end{equation}

We can restore specific values of $\gamma$ for different models of matter fields. For example, we have $\gamma = \frac{5}{16}$ for dust ($w = 0$),
$\gamma = 0$ for radiation ($w = \frac{1}{3}$), $\gamma = - \frac{1}{4}$ for stiff matter ($w = 1$), and so on. If $w = - 1$, then $\gamma = 2$
corresponding to the exponentially expanding universe (see Eq.~(\ref{17}). Recall that the model of matter with the equation of state $p = w \rho$
is completely identical to the model considered above in this Section, if we put $\alpha = 2/[3 (1 + w)]$.

It is easy to see that the values $\alpha \ll 1$ can be obtained, if the equation of state parameter $w \gg 1$, which according to Eq.~(\ref{181}) yields
$\gamma \gg 1$. The models with the large values of the EoS parameter are known. They are considered, for instance, in the context of the 
Ekpyrosis scenario (see, e.g., Refs.~\cite{Kh01,Kh02,St19}). However, in our approach, we are dealing with a period of ultra-slow expansion rather than 
contraction.

\section{Stiff matter}\label{sec:5}
The dependence $\rho \sim a^{-6}$ appears also in the case, when the universe is filled with a cosmological scalar field whose energy density is dominated by 
its kinetic term. It was found that physical models introducing EDE that modifies the expansion history of the universe before the 
recombination period and then decays faster than radiation can resolve the Hubble tension \cite{R22,R21,V21,V22,Ka1806,Ka1811}. Many of the EDE models 
that have been explored postulate an oscillating scalar field potential. Concerning the approach based on quantum cosmology, it has the advantage, since a 
desirable dependence of the energy density on $a$ arises naturally and does not require any model assumptions.

From the point of view of the standard big bang model, the quantum addition to the energy density (\ref{18}) can be considered as a stiff matter (fluid)
with the dynamical coupling constant $\gamma$ that changes during the evolution of the universe.
Taking into account its application for solving different cosmological puzzles, it is of interest to consider the appearance of the aforementioned dependence of 
the energy density on the scale factor without involving quantum cosmology.

As is well known, the origin of dependence $a^{-6}$ can be observed in general relativity.
Introducing the EoS parameter for a single matter component, $w = p / \rho = const$, from energy conservation, one gets
$\rho \sim a^{- 3(1 + w)}$. Then $\rho \sim a^{- 6}$ attributed to stiff matter component corresponds to $w = 1$.

The cosmological model in which the universe near the cosmological singularity is assumed to be filled with a gas of cold baryons with a stiff equation of state 
has been first introduced in Ref.~\cite{Zel72}. It is supposed that a stiff matter era preceded the radiation era, the dust matter era, and the dark energy era. 
This stiff matter era also occurs in certain cosmological models where dark matter is made of relativistic self-gravitating Bose-Einstein condensates 
\cite{Cha15}. The energy density of the stiff matter can be positive or negative depending on the nature of the self-interaction of the bosons leading to different 
scenarios of the evolution of the universe.

The effects of a stiff matter era preceding recombination on the energy spectrum of the primordial gravitational waves can be examined \cite{Oik23}.
A stiff pre-recombination era, also referred to as the kination era, can be caused by various mechanisms. It was argued that if the universe underwent an era of 
``kination'', with the energy density dominated by the kinetic energy of a classical homogeneous scalar field, then the energy density falls very rapidly as 
$a^{-6}$, faster than radiation, which can lead to interesting physical phenomena \cite{Co22}. The origins of the kination era can be interpreted, in particular, 
within the framework of the axion kination approach \cite{Co20}.  In the case of the electroweak symmetry breaking, the axion potential acquires a new 
minimum, and the axion rolls swiftly to this minimum, experiencing a short kination epoch, where its energy density redshifts as $a^{-6}$ \cite{Oik23a}.

One can observe the origin of stiff matter in a two scalar field model that incorporates non-Riemannian measures of integration and involves an additional 
$R^{2}$--term as well as scalar matter field potentials of appropriate form so that the action is invariant under global Weyl-scale symmetry \cite{Gue23,Gue14}. 
The model allows for a unified description of both early universe inflation as well as of late dark energy epoch. It possesses a non-singular emergent universe 
solution which describes an initial phase of evolution that precedes the inflationary era. The aim of this theory is to provide an explanation for generation of 
dark energy and dark matter in the late universe, which appear as a consequence of the K-essence produced by the additional kinetic terms and 
$R^{2}$--terms introduced into the action. The contributions to the dark energy come from an effective scalar field potential and from a K-essence background 
configuration. Furthermore, for the present universe, one has a dark matter component and an additional stiff matter component which originate from the 
perturbation of the background K-essence configuration. It is worth mentioning that in the emergent universe scenario, which can be seen as a modern version 
and extension of the Eddington universe, the inflationary era emerges from a static state, where the Hubble expansion rate becomes zero, $H = 0$ 
\cite{Gue23,Ell04}. Similarly, in our approach (see Sect.~\ref{sec:4}), the state in which the contribution of the stiff matter becomes important also arises from 
an almost static state.

The effect of a non-interacting fluid with the equation of state of the stiff matter, introduced into the cosmological model, is dynamically equivalent to the action 
of perfect fluid with spin known as the Weyssenhoff fluid \cite{Wey47,Szy04}. In the framework of Einstein--Cartan theory of gravity, the antisymmetric part of 
the affine connection coefficients (torsion) becomes an independent dynamic variable which can be associated with the spin density of matter $s_{\mu \nu}$ in 
the universe. The contribution $\sigma^{2}$ of the macroscopic spin of the fluid to the total energy density $\rho_{\mbox{\tiny eff}}$ is negative, 
$\rho_{\mbox{\tiny eff}} = \rho - \sigma^{2}$, and it is given by the square of the spin density, $\sigma^{2} = \frac{1}{2} s_{\mu \nu} s^{\mu \nu} \sim a^{-6}$.
The equations of the Einstein--Cartan theory of gravity with torsion for the Friedmann--Robertson--Walker universe can be recognized as the 
Eistein--Friedmann equations with quantum corrections \cite{K13}. On the other hand, the addition of stiff matter to the cosmological model produces the effect formally equivalent to the brane effects with dust on a brane with negative tension \cite{Szy02}.

\section{Numerical estimations}\label{sec:6}
We have shown that in the model of a single dominant matter field, the factor $\gamma$ in (\ref{18}) can take different values at different stages of the 
universe's evolution, when different matter-energy components dominate. As was pointed out, the factor $\gamma = 0$ and no additional quantum energy 
density $\rho_{B}$ is produced in the early stages of the evolution of the universe, when a dominant matter field is radiation ($\alpha = \frac{1}{2}$). When the 
contribution of non-relativistic matter component and the curvature term become comparable to the contribution of relativistic matter, the quantum addition to 
energy density begins to be generated, while remaining small, since the factor $\gamma$ takes values close to unity. For further numerical estimation, we can 
assume that this occurs prior, but close to the epoch of matter-radiation equality (approximately at a redshift $z_{eq} \approx 10^{3.5}$, when the scale factor 
$a_{eq} \approx 5.4 \times 10^{57}$, setting the scale factor at the present epoch equal to the current Hubble radius 
$a_{0} = c H_{0\, \mbox{\tiny SNe}}^{-1}/l_{p} = 1.7 \times 10^{61}$).

In the following, it is convenient to accept that the universe is closed and to work with a space of positive curvature, as in the Lema\^{i}tre and 
Eddington--Lema\^{i}tre models. Despite the fact that there is a strong research community inclination toward a flat universe, it was shown that
the Planck power spectra had a moderate preference for closed universes \cite{Ha19,Va20}. In this case, we can consider the universe going through a 
``coasting period'' during which the scale factor remains close to some value. Then the effective parameter $\alpha$ approaches zero (and 
accordingly, the EoS parameter $w$ goes to infinity), so the factor $\gamma$ takes on large values. The growth of the factor $\gamma$ 
naturally ceases when the contribution from the corresponding additional energy density $\rho_{B}$ of the quantum nature becomes commensurate with the 
curvature term, and the universe exits the ``coasting period''. It is reasonable to expect that the ``coasting period'' comes along with the epoch of 
matter-radiation equality $z_{c} \approx z_{eq}$, and finishes after its end, before recombination (CMB photons last scatter at the redshift $
z_{ls} \approx 1.1 \times 10^{3}$, when the scale factor $a_{ls} \approx 1.5 \times 10^{58}$).

The additional quantum energy density contributes a fraction of the energy density of the universe,
\begin{equation}\label{182}
f_{B} (a) = \frac{\rho_{B}(a)}{\rho(a)}.
\end{equation}
Early-time solutions to the Hubble tension postulate an increase in the energy density due to some extra component with a fractional contribution of 
$\sim 10\%$ to the total energy density briefly before recombination \cite{KR22}. In our approach, a quantum addition to the energy density can provide the 
required increase in the energy density in the specified epoch. Considering $f_{B} (a_{c})$ as the normalization parameter that brings the expected 
contribution to the total energy density and resolves the Hubble tension \cite{KR22,L19,L22}
\begin{equation}\label{183}
f_{B} (a_{c}) \approx 0.1,
\end{equation}
we can estimate the value of the factor $\gamma$,
\begin{equation}\label{184}
\gamma \approx 3 \times 10^{232}.
\end{equation}
On the other hand, the growth of the factor $\gamma$ is determined by the period ending when the contribution from the energy density $\rho_{B}$ of the 
quantum nature becomes comparable to the curvature term. For the factor $\gamma$ (\ref{184}), this occurs earlier then the scale factor reaches the value $a_{B} \approx 1.3 \times 10^{58}$, i.e., before recombination.

Note that even for a large value of the factor $\gamma$, the additional energy density $\rho_{B}$ of the quantum nature provides the desired rapid decay
with scale factor $a$ as $\rho_{B} \sim a^{-6}$.

In contrast to various EDE models (which encounter the so-called coincidence problem \cite{L22,Ka16}), the approach we propose immediately makes it clear 
why the epoch of influence of the additional energy density component corresponds to the epoch of matter-energy equality, preceding recombination. It is in 
this epoch that an additional energy density of quantum nature is generated, and its desired fractional contribution to the total energy density is achieved not by 
a simple fitting of a free parameter, but by the universe's exit from the ``coasting period'' during which the factor $\gamma$ acquires a corresponding large 
value.

\section{Conclusion}\label{sec:7}
In the quantum cosmological approach studied in this note, there naturally arises an additional energy density (\ref{18}) that dilutes away faster than radiation.
Such a behavior is stipulated by a well-known Bohm potential (\ref{7}). This energy density component that has a quantum origin acts similarly to
a stiff matter component introduced into the cosmological model. It may have the same effect as
EDE modifying the early expansion rate before recombination while leaving the late evolution of the universe unchanged. For this reason, the model
based on quantum cosmology can be considered as a candidate for solving the Hubble tension.

There is another implication from the calculations given above, which are supported by the equations of quantum cosmology. Despite the 
generally accepted point of view that in a universe that has reached the pre-recombination epoch, one cannot expect quantum effects to manifest themselves, 
their influence can be revealed by comparing cosmological parameters extracted from the data of astronomical observations by using different model methods.

The generalized Einstein--Friedmann equation (\ref{14}) allows us to supplement the standard $\Lambda$CDM model with a new term of quantum nature:
\begin{eqnarray}\label{19}
H^{2}(z) = H_{0}^{2} \left [\Omega_{m 0} (1 + z)^{3} + \Omega_{dm 0}(z) (1 + z)^{3} + \Omega_{\gamma 0} (1 + z)^{4}  \right. \\ \nonumber
+ \left. \Omega_{\Lambda 0} + \Omega_{curv 0} (1 + z)^{2} + \Omega_{B 0} (1 + z)^{6}   \right ],
\end{eqnarray}
where $\Omega$'s are the energy densities in units of critical density, $z$ is a redshift, $\Omega_{B} (z) = \Omega_{B 0} (1 + z)^{6}$ is the quantum energy 
density corresponding to $\rho_{B}$ (\ref{16}). The energy density of dark matter $\Omega_{dm 0}$ is a function of $z$, since it can contain both 
non-relativistic and relativistic matter of any nature. From Eq.~(\ref{19}), it follows the constraint
\begin{equation}\label{20}
\Omega_{\Lambda 0} = 1 - \Omega_{m 0} - \Omega_{\gamma 0} - \Omega_{dm 0}(0) - \Omega_{curv 0} - \Omega_{B 0}.
\end{equation}
For $\Omega_{curv} = 0$ and $\Omega_{B} = 0$, one has the standard model of spatially flat universe.

One can estimate the contribution from an additional energy density $\Omega_{B 0}$ at the present epoch. Choosing
a critical energy density corresponding to the Hubble expansion rate $H_{0\, \mbox{\tiny SNe}} = 73.04$ km/s/Mpc, we find
\begin{equation}\label{201}
\Omega_{B 0} = \gamma \times 0.13 \times 10^{-244}.
\end{equation}
For the factor $\gamma$ (\ref{184}), it gives $\Omega_{B 0} = 0.39 \times 10^{-12}$. In other words, the late evolution of the universe remains unchanged
and this explains why the contribution of this additional energy density is not easy to observe with astrophysical surveys.

We can find out at a glance at what redshift $z_{B}$ the contribution of the additional energy density $\Omega_{B}$ will be comparable to that of non-relativistic matter (pressureless matter parameter). Taking $\Omega_{m 0} + \Omega_{dm 0}(0) = 0.315$ \cite{PDG}, we find that $z_{B} \sim 10^{4}$.

\section*{Acknowledgements}
The present work was partially supported by The National Academy of
Sciences of Ukraine (Projects No.~0121U109612 and  No.~0122U000886) and by a grant from the Simons Foundation (Grant Number 1030283, VEK, VVK).

\end{document}